\documentclass[twocolumn,letterpaper,showpacs,preprintnumbers,amsmath,prb]{revtex4}
\usepackage{graphicx}
\usepackage{color}
\usepackage{verbatim}
\usepackage{hyperref}

\def\k{\mathbf{k}}
\def\q{\mathbf{q}}

\def\d{\mathrm{d}}
\def\D{\mathcal{D}}
\def\Tr{\operatorname{Tr}}
\def\Re{\operatorname{Re}}
\def\oo{\infty}
\def\eps{\varepsilon}

\begin{document}
\title{Formation of Electronic Nematic Phase in Interacting Systems}
\author{Igor Khavkine$^1$, Chung-Hou Chung$^1$, Vadim Oganesyan$^2$, 
Hae-Young Kee$^1$}
\affiliation{$^1$ Department of Physics, University of Toronto,
Toronto, Ontario  M5S 1A7,  Canada \\
$^2$ Department of Physics, Princeton University,
Princeton, New Jersey 08544, USA }

\begin{abstract}
We study the formation of an electronic nematic phase characterized by a
broken point-group symmetry in interacting fermion systems within the
weak coupling theory. As a function of interaction strength and chemical
potential, the phase transition between the isotropic Fermi liquid and
nematic phase is first order at zero temperature
and becomes second order at a finite temperature. The
transition is present for all typical, including quasi-2D,
electronic dispersions on the square lattice and takes place for
arbitrarily small interaction when at van~Hove filling, thus
suppressing the Lifshitz transition. In connection with the formation
of the nematic phase, we discuss the origin of the first order
transition and competition with other broken symmetry states.
\end{abstract}
\maketitle

\section{Introduction\label{sec:intro}}

Recently, there have been reports of experimental evidence of
inhomogeneous and/or anisotropic quantum ground states
in strongly correlated
systems.\cite{cheong,eisenstein1,eisenstein2,eisenstein3,ando-segawa}
The inhomogeneous phase---dubbed the stripe phase---%
was observed in high temperature cuprates via elastic
neutron scattering experiments.\cite{tranquada}
Theoretical studies of the inhomogeneous and anisotropic
quantum ground states in connection with generic phases of a doped Mott
insulator have also been carried out.\cite{kivelson,kivelson-bis}
It was suggested that as quantum fluctuation, induced by hole doping,
increases, the Mott insulator turns into a smectic---stripe---phase,
and a further increase of quantum fluctuation will turn a smectic to
a nematic phase.\cite{kivelson,kivelson-bis}
The electronic smectic and nematic phases 
are characterized by broken translational (in one direction) 
and rotational symmetries, in analogy with
classical liquid crystals.\cite{de_gennes}

The electronic nematic phase which breaks the point-group symmetry
on a square lattice was studied in the extended Hubbard model.
The instability of Fermi liquid towards  possible ordered phases
including the nematic---called the Pomeranchuk
instability\cite{pomeranchuk}---%
were investigated, and it was shown that the nematic phase
is the leading instability within some range of the parameter
space.\cite{metzner,wegner,kampf} 
Within the weak coupling theory,
the effective Hamiltonian of the quadrupole-quadrupole density interaction 
for the nematic phase was also constructed,
where the expectation value of the quadrupole density  
is the order parameter of the nematic phase.\cite{oganesyan}
The study of the nematic phase and possible superconductivity
in the continuum model within the weak coupling approach showed
the non-Fermi liquid behavior in 
the single particle scattering rate\cite{oganesyan} 
and an exotic superconducting pairing symmetry\cite{kim}
via the coupling to the collective modes.
Possible probes of the nematic phase were also discussed.\cite{kee,kivelson2}

The nature of the quantum phase transition between the nematic phase and
isotropic Fermi liquid as a function of chemical potential, for a
particular set of parameters (interaction strength $F_2$ and next
nearest neighbor hopping $t'$), was recently studied
numerically.\cite{kee2} This  computation showed that the transition is
generically first order. On the other hand, Ref.~\onlinecite{metznerrohe}
found anisotropic non-Fermi-liquid behavior at the isotropic-nematic
quantum critical point---second order transition---in the presence of a
lattice. Therefore it is important to investigate whether the nature of
isotropic-nematic transition depends on the type of electron dispersion,
the interaction strength, or the temperature.

In this paper, we investigate the behavior of the free energy and
density analytically, as a function of chemical potential and
interaction strength. We also extend the analysis to finite temperature
and quasi-2D electron dispersion. At zero temperature, we show that the
nematic transition is first order as a function of the interaction
strength and chemical potential for all typical 2D electronic
dispersions since they possess van~Hove singularities. The transition
changes to a continuous one at a finite temperature, but is not strongly
affected by small dispersion in the third direction.

We also find that the transition takes place for arbitrarily small
attractive interaction at the van Hove band filling.
In the absence of interaction, at the van~Hove filling, the Fermi
surface changes topology from electron- to hole-like. This transition,
first studied by Lifshitz\cite{lifshitz,lifshitz-ru}, causes
singularities in thermodynamic quantities such as compressibility, due
to van~Hove singularities, but is not accompanied by any broken symmetry.
This transition was recently revisited in Ref.~\onlinecite{wen} as an
example of phase transition between different quantum orders which are
not classified by broken symmetry.
Our findings show that the Lifshitz transition does not take place
because the van~Hove singularity is avoided due to a sudden change of
the Fermi surface topology. This has implications for earlier studies of
broken symmetry states, such as density waves. These
studies\cite{schulz,lederer,furukawa,kampf,honerkamp,metzner}
were based on the existence of the van~Hove singularity, which lead
to a divergence of the relevant susceptibility indicating a transition
to an ordered phase. Our results suggest that these ordered phases
may be preempted by the first order transition into the nematic phase.

The paper is organized as follows.
We describe the effective model Hamiltonian for the nematic
in section \ref{sec:model}. The mean field analysis at zero temperature
and finite temperature is given in  section \ref{sec:mfa}.
A possible connection to other competing
instabilities is discussed in section \ref{sec:summary}.
We also provide the summary of our findings in the last section,
\ref{sec:summary}.

\section{Model\label{sec:model}}
\subsection{Hamiltonian}
Our choice of model Hamiltonian is largely motivated by symmetry
considerations and by the philosophy of the weak-coupling BCS theory.
The interaction is chosen with foresight toward the
mean-field analysis and with inspiration from classical liquid crystal
theory.
In a liquid crystal, each rod-like molecule defines a direction in
space, and the order parameter for the nematic phase is equally
sensitive to their alignment as well as anti-alignment. It is a
quadrupole (second order symmetric traceless) tensor built from these
directions. In two dimensions, it changes sign under a rotation by
$90^\circ$ and is invariant under a rotation by $180^\circ$.
For an electron gas, we can construct a similar order parameter from the
momenta of electrons, the quadrupole density $\hat{Q}_{ij} =
\hat{p}_i\hat{p}_j-\frac{1}{2}\hat{p}^2\delta_{ij}$.  The interaction
between quadrupole densities (suitably quantized and discretized) is
made attractive to favor the alignment or anti-alignment of electron
momenta, i.e.\ formation of the nematic phase.

The precise Hamiltonian under consideration is written as
\begin{equation}\label{H-def}
        H = \sum_{\k} (\eps_\k-\mu) c^\dagger_\k c_\k
                + \sum_{\q} F_2(\q)
                \Tr\left[\hat{Q}^\dagger(\q)\hat{Q}(\q)\right],
\end{equation}
where $\eps_\k$ is the single-particle dispersion, $F_2(\q)$ is the
inter-electron interaction strength, and $\hat{Q}(\q)$ is the the lattice
analog of the quadrupole density tensor constructed in two dimensional
square lattice from particle momenta given by
\begin{equation}
        \hat{Q}(\q) = \! \frac{1}{N} \sum_{\k} c^\dagger_{\k+\q}
        \begin{pmatrix}
                \cos k_y - \cos k_x & \sin k_x \sin k_y \\
                \sin k_x \sin k_y  & \cos k_x - \cos k_y
        \end{pmatrix} c_{\k}.
\end{equation}
The dispersion is that of a next-nearest-neighbor tight binding model on a
square lattice $\eps_\k = -2t(\cos k_x + \cos k_y)-4t'\cos k_x \cos k_y$.
While $F_2(\q)$ is an arbitrary short ranged interaction, for
example of the (two dimensional) Yukawa form,
\begin{equation}\label{F2q}
        F_2(\q) = -\frac{F_2}{2}\frac{2\pi\kappa^2}{(1+\kappa^2 \q^2)^{3/2}}.
\end{equation}
We shall only be interested in the strength of the interaction $F_2$,
which is assumed to be 
positive.

\begin{figure}
\begin{center}
	\includegraphics{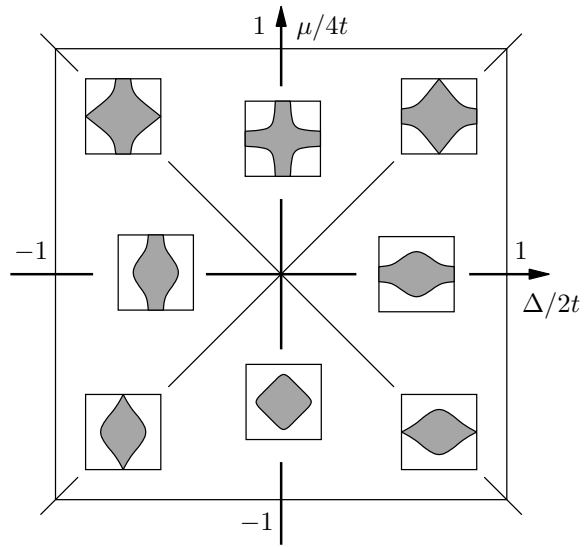}
\end{center}
        \caption{Shapes of the Fermi surface (FS) for different values of $\mu$ and
                $\Delta$. Lines $(\mu/2)^2=\Delta^2$ mark the van Hove singularities in the
                DOS which correspond to the FS touching the boundary of the
                Brillouin zone.\label{dos-phdiag}}
\end{figure}
\section{Mean-field analysis\label{sec:mfa}}
To decouple the quartic interaction in Eq.~\eqref{H-def}, we concentrate
on the $\q=0$ interactions (which is equivalent to letting
$F_2(\q)\to -\frac{1}{2}F_2\delta_{\q,0}$ as $\kappa\to\oo$ in Eq.~\eqref{F2q})
and define two order parameters,
\begin{align}
        \Delta &= F_2 \langle \hat{Q}_{xx}(0) \rangle &&\text{and}&
        \Delta' &= F_2 \langle \hat{Q}_{xy}(0) \rangle,
\end{align}
where the expectation value $\langle c^\dagger_\k c_\k \rangle$ is
replaced by the Fermi distribution $n_F(\eps_\k-\mu)$, and $\eps_\k$ is
the the renormalized single-particle dispersion relation
\begin{multline}\label{xik}
        \eps_\k = -2t(\cos k_x + \cos k_y) - 4t'\cos k_x \cos k_y \\
                + \Delta(\cos k_x - \cos k_y)
                - \Delta' \sin k_x \sin k_y.
\end{multline}
A finite $t'$ does not qualitatively change the physics of the
model. For simplicity, the value $t'=0$ was used in the $T=0$ calculations,
but a nonzero $t'$ was used at finite temperature.

A non-zero expectation value of $\hat{Q}_{ij}$ indicates that the
rotational (lattice point-group) symmetry has been broken and a
preferred direction for electron momenta has been selected.  In
particular, a non-zero expectation value of $\hat{Q}_{xy}$ indicates
that this direction is not parallel to either of the crystal axes.  The
profile of both the interaction and the bare dispersion favor alignment
along the crystal axes, as long as the coupling constants for diagonal
$\hat{Q}_{xx}$ and off-diagonal $\hat{Q}_{xy}$ elements are the same,
$F_2^{xx}=F_2^{xy}$. Hence we expect $\Delta'$ to vanish. This
observation has been confirmed by numerical calculations.\cite{kee2}

In general, unlike in the continuum model\cite{oganesyan} with full
rotational symmetry, the coupling constants $F_2^{xx}$ and $F_2^{xy}$
can be different.  In such cases, a finite $\Delta'$ is possible.  In
fact broken symmetry states with oblique alignment, such as diagonal
stripes, have been discussed in Ref.~\onlinecite{fujita-etal}.

\subsection{Free energy}
The resulting mean-field grand-canonical free energy density is given by
\begin{equation}\label{FE-def}
        F(\mu,\Delta) = \frac{1}{F_2}\frac{\Delta^2}{2} + F_0(\mu,\Delta),
\end{equation}
where $F_0$ is
\begin{equation}
        F_0(\mu,\Delta) = -\frac{1}{\beta}\int\d{\eps}\, \D(\eps)
                \ln(1+e^{-\beta(\eps-\mu)}),
\end{equation}
with $\D(\eps)$ being the density of states (DOS) of the single-particle
dispersion \eqref{xik}. Its exact expression is (see Appendix for
details and definitions)
\begin{equation}\label{DOS-def}
        \D(\eps) = N_0\Re\left\{ \frac{1}{\sqrt{1-(\frac{\eps}{4t})^2}}
                K\left(1-\frac{\Delta^2-(\frac{\eps}{2})^2}{(2t)^2-(\frac{\eps}{2})^2}\right)
                \right\},
\end{equation}
where $N_0= 1/(2t\pi^2)$. For $|\eps|>4t$, $\D(\eps)$ vanishes.  Note
the logarithmic van~Hove (vH) singularities in the DOS at $(\eps/2)^2 =
\Delta^2$. These singularities occur when the constant energy contour
(Fermi surface) meets the boundary of the first Brillouin zone as shown
in Fig.~\ref{dos-phdiag}.
Due to these remnants of the vH singularity of the non-interacting
system, the free energy $F(\mu,\Delta)$ will exhibit non-analytic
behavior at $(\mu/2)^2=\Delta^2$.  This behavior is to be contrasted
with the free energy in the presence of density wave order parameters,
where all singularities are smoothed. Ultimately, persistence of
non-analyticities in $F(\mu,\Delta)$ leads to the first order
isotropic-nematic quantum phase transition.

\subsection{Zero temperature}
First we analyze the free energy density in the limit of zero temperature
($\beta\to\oo$). In this limit
\begin{equation}
        F_0(\mu,\Delta) = (E-\mu n),
\end{equation}
where $E$ and $n$ are the energy and particle density per unit cell.

The energy integral can be evaluated exactly, but the density integral
cannot. The combined expression for the free energy (for $\Delta,
\frac{\mu}{2} < 2t$, and neglecting terms independent of $\Delta$) is
\begin{multline}\label{FE-expr}
        F = \left(\frac{1}{F_2}+2N_0\right)\frac{\Delta^2}{2} \\
                {}+ N_0\left[
                        \left(\Delta+\frac{\mu}{2}\right)^2
                               \ln\left|\frac{\Delta+\frac{\mu}{2}}{4}\right| + (\mu\to-\mu)
                \right],
\end{multline}
where for brevity all energy quantities are in units of $2t$.
Any results extracted from this expression are valid up to
quadratic order in $\mu$ and $\Delta$. 
The details of the calculations are outlined in the Appendix.

\begin{figure}
\begin{center}
	\includegraphics{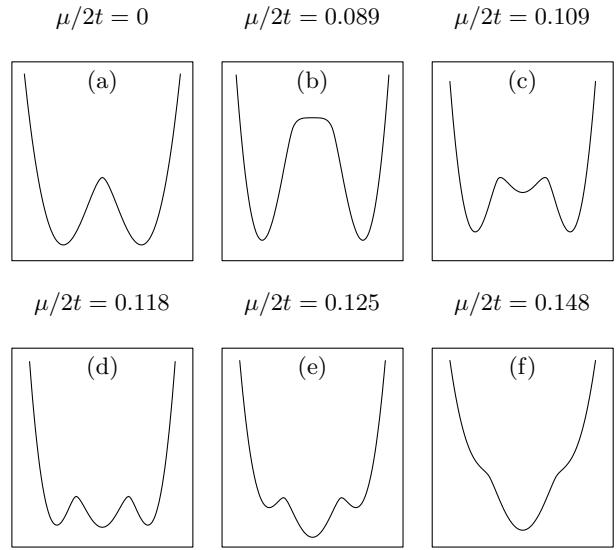}
	\caption{Plots of free energy as a function of $\Delta$ for different
	values of the chemical potential ($F_2 N_0=0.1$) centered at
	$\Delta=0$; (b) $\mu=\mu_*$, (d) $\mu=\mu_c$ (see text).\label{FE-plots}}
\end{center}
\end{figure}
The free energy for different chemical potentials is plotted
as a function of $\Delta$ in Fig.~\ref{FE-plots}. As the chemical
potential decreases, the free energy develops local minima
at finite $\Delta$ (the nematic phase), which then become the global
minima for $|\mu|<\mu_c$. It is clear that the transition between
isotropic and nematic phases is first order.

\begin{figure}
\begin{center}
	\includegraphics{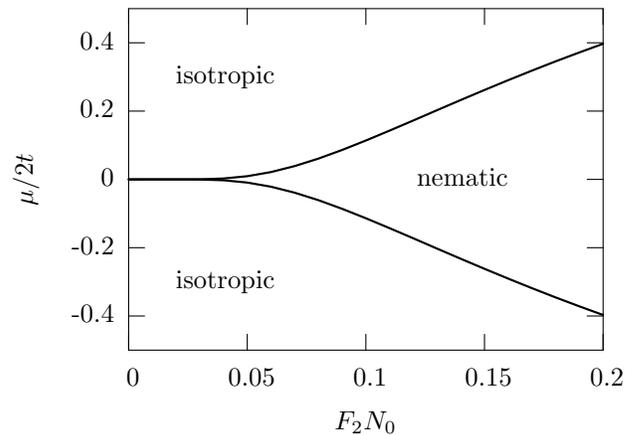}
\end{center}
        \caption{Phase diagram in the $\mu$-$F_2$ parameter space. Solid
				curve, given by $\mu_c$ \eqref{mu-critical}, is the line of the
				first order transition.\label{mu_vs_F2}}
\end{figure}
\begin{figure}
\begin{center}
	\includegraphics{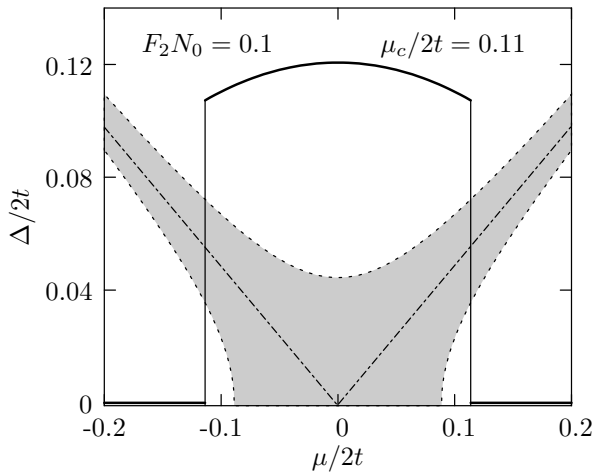}
\end{center}
        \caption{Equilibrium value of the order parameter $\Delta$.
				The shaded region indicates negative curvature of the free
				energy. Lines $\Delta^2=(\frac{\mu}{2})^2$ are marked by
				dot-dashed lines.
				\label{D_vs_mu}}
\end{figure}
From Eq.~\eqref{FE-expr}, one can also show that $\Delta=0$ is always
an extremum of the free energy. However, no local minimum of the free
energy can be located in the region $|(\frac{\mu}{2})^2-\Delta^2|
< (\frac{\mu_*}{2})^2$, where
\begin{equation}\label{mu-stability}
        \mu_*/2t = 1.08 e^{-\frac{1}{4F_2 N_0}},
\end{equation}
this is the limit of metastability of the isotropic phase ($\Delta=0$)
which becomes unstable for $|\mu|<\mu_*$.
Hence, $\Delta$ must have a finite equilibrium value in this
region (nematic phase). The phase transition actually takes place
slightly outside this region at $|\mu|=\mu_c$,
\begin{equation}\label{mu-critical}
        \mu_c/2t = 1.39 e^{-\frac{1}{4F_2 N_0}}.
\end{equation}
We stress again that $\mu_c > \mu_*$, which indicates that the
nematic transition takes precedence over the Pomeranchuk instability
(divergence of susceptibility).
In Fig.~\ref{mu_vs_F2}, we show the line of the first order phase
transition in the parameter space of interaction strength, $F_2$,
and chemical potential.

The nontrivial local minima are located at
\begin{equation}\label{Delta-eql}
        \pm \Delta/2t = 1.47e^{-\frac{1}{4F_2 N_0}}
                - 1.36 e^{\frac{1}{4F_2 N_0}} \left(\frac{\mu}{2t}\right)^2.
\end{equation}
Eq.~\ref{Delta-eql} is valid for $|\mu|<\mu_c$.
The Fig.~\ref{D_vs_mu} shows the behavior of the order parameter.  The
order parameter jump  and the width of the nematic window decrease
exponentially as  the coupling strength $F_2$ goes to $0$.
The unstable region $|\Delta^2-(\frac{\mu}{2})^2|<(\frac{\mu_*}{2})^2$
is shaded in Fig.~\ref{D_vs_mu}. A local minimum
of the free energy can only be found outside this area, which explains
the discontinuous character of the phase transition, since the
equilibrium value of the order parameter must jump to avoid it.
The dot-dashed lines, $(\frac{\mu}{2})^2=\Delta^2$, are where the
change in topology of the Fermi surface takes place, the Lifshitz
transition. As shown in Fig.~\ref{D_vs_mu}, these lines are embedded in
the unstable region.
Therefore, our
results indicate that, in the presence of interaction, the Lifshitz
transition  is not realized due to the formation of the nematic phase.

\begin{figure}
\begin{center}
	\includegraphics{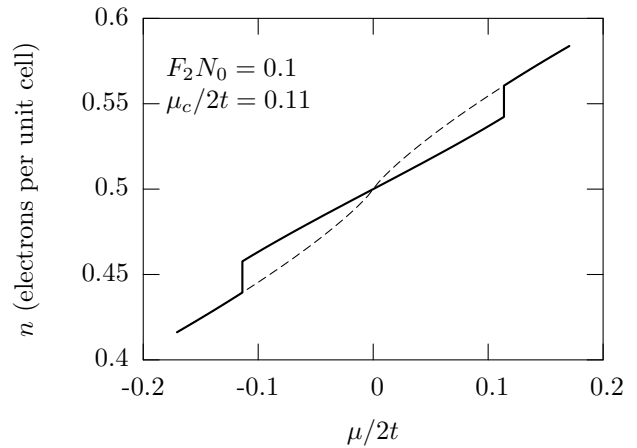}
\end{center}
        \caption{The electron density per unit cell. The jump in the
        density is a signature of a first order phase transition.\label{n_vs_mu}}
\end{figure}

The density as a function of $\mu$ is shown in Fig.~\ref{n_vs_mu}.
Notice that the dotted line is the density in the absence of nematic
order.  Its derivative at $\mu=0$ is singular, which signals the
Lifshitz transition. However, once the nematic order sets in, the
density at half filling becomes smooth, but shows a discontinuity at
$\mu_c$.

\subsection{Finite temperature}
\begin{figure}
	\includegraphics{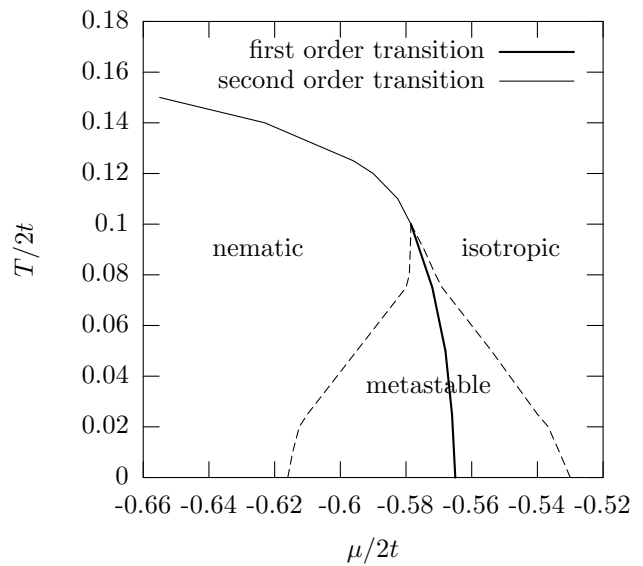}
\caption{The finite temperature phase diagram for the
        isotropic-nematic transition. The first order transition persists at
        low temperatures. However, the metastable region gets smaller and
        smaller with increasing temperature. Finally, around $T/2t = 0.1$,
        the phase transition becomes continuous.\label{finite-T}}
\end{figure}
To investigate the robustness of the first order iso\-tro\-pic-nematic
transition, we study the transition at finite temperature and with a
finite inter-plane hopping term $t_z$. These calculations are performed
numerically using the same technique as in Ref.~\onlinecite{kee2}.  The first
order transition is robust against a small $t_z=0.1t$ term, but changes to a
second order one at a finite temperature.

The phase diagram for finite temperatures is shown in
Fig.~\ref{finite-T} for a fixed $F_2 N_0 = 0.11$ and a finite
next nearest neighbor hopping $t'$ ($t' =-0.4 t$). The full width of
the nematic window is from $\mu/2t=-0.94$ to $-0.57$.\cite{kee2}
It is worth noting that the first order transition does not alter
qualitatively with a finite $t'$.

A negative (positive) $t'$ shifts the window of the nematic phase
to the hole- (electron-) doped side.
At low temperatures the transition is still first
order surrounded by a metastable region (which indicates the presence
of unstable local minima). However, at about $T/2t = 0.1$, the
metastable region disappears and the transition becomes continuous.

The discontinuity in the isotropic-nematic transition at zero
temperature can be traced to
the presence of the lattice, which dictates the form of the dispersion
relation and the presence of the van Hove singularity, and the sharpness
of Fermi distribution.
With increasing temperature, the thermal fluctuation will
smear the sharpness of the Fermi distribution function, which 
%
results in a smaller jump in the
order parameter and finally in a continuous transition. This 
expectation is confirmed by the numerics presented here.

\section{Discussion and Summary\label{sec:summary}}
There has been a number of one-loop renormalization group studies for
the Hubbard model taking into account the Fermi surface only at the
saddle points, $(\pm \pi,0)$ and $(0, \pm \pi)$, namely the two patch model at
van Hove filling.  These studies showed that there are antiferromagnetic
and $d$-wave pairing instabilities in Hubbard
model.\cite{schulz,lederer,dylo} Recently the two patch model was
revisited, and truncation of the Fermi surface near the saddle points
was suggested.\cite{furukawa}

On the other hand, the instability of the Fermi li\-quid towards the formation
of the nematic phase---Pomeranchuk instability---with other competing
orders were also recently investigated using different methods in the
extended Hubbard model\cite{kampf,metzner,wegner,honerkamp} and the
$t$-$J$ model.\cite{yamase-kohno}
%
In Ref.~\onlinecite{metzner}, the authors noticed that the nematic instability
is driven by the attractive (repulsive) interaction between
electrons in the same (different) patches via forward scatterings
in the Hubbard model.
This finding is consistent with our effective Hamiltonian, where 
$-F_2(\cos{k_x}-\cos{k_y})(\cos{k^{\prime}_x} -\cos{k^{\prime}_y})$ suggests
an attractive interaction between electrons near  
$(\pm \pi,0)$ and $ (\pm \pi, 0)$, and a repulsive one between
$(\pm \pi,0)$ and $ (0, \pm \pi)$. 
While the understanding of the effective interaction
for the nematic phase from the microscopic Hamiltonian 
is still missing, it suggests that
the effective nematic interaction is hidden in the extended Hubbard model.

Our discovery of a strong tendency toward the nematic phase near van
Hove filling suggests that the two patch model should be revisited.
Since the nematic order occurs for extremely small interaction near van
Hove filling, the existence of saddle points itself should be addressed.
Since the Fermi surface topology ``suddenly'' changes from closed to open
as indicated in our result, we speculate that the instability towards
competing orders such as antiferromagnetic and charge density wave
(which are sensitive to the topology of the Fermi surface and enhanced
by van Hove singularity) would be suppressed by the formation of the
nematic phase.

In summary, we have investigated a model Hamiltonian exhibiting the
nematic phase. At zero temperature the isotropic-nematic transition
takes place for arbitrarily small coupling at van Hove band filling.
Away from the van Hove filling, a finite minimum interaction is required
to stabilize the nematic phase.  The phase transition is first order as
a function of interaction strength (chemical potential) for a fixed
chemical potential (interaction strength) as shown in
Fig.~\ref{mu_vs_F2}.  The strong tendency toward the nematic phase for
an arbitrary small interaction at van Hove filling suggests that the
Lifshitz transition is suppressed in the presence of interactions.  At a
finite temperature the transition becomes second order, while it remains
first order for a quasi-2D dispersion.

\begin{acknowledgments}
This work was supported by NSERC of Canada (IK, CHC, HYK), Canada Research
Chair (HYK), Canadian Institute for Advanced Research (HYK), Alfred P.\
Sloan Research Fellowship (HYK), and Emerging Material Knowledge program
funded by Materials and Manufacturing Ontario (HYK).
VO thanks W.~Metzner and D.~Huse for discussions. VO is supported by
grants from NSF (DMR 99-78074 and DMR 02-13706), and from the David and
Lucille Packard foundation.
\end{acknowledgments}

\appendix
\section{Derivation of free energy\label{sec:fe-appendix}}
The Density of states is defined and evaluated as
\begin{multline}
	\D(\eps) = \frac{1}{N}\sum_{\k} \delta(\eps-\eps_k) \\
	= N_0\Re\left\{\frac{1}{\sqrt{1-(\frac{\eps}{4t})^2}}
		K\left(1-\frac{\Delta^2-(\frac{\eps}{2})^2}
			{(2t)^2-(\frac{\eps}{2})^2}\right)\right\},
\end{multline}
where $K(m)$ is the Complete Elliptic Integral of the First
Kind,\cite{ASellint} and $N_0=1/(2\pi^2 t)$.
The function $K(m)$ has a logarithmic singularity at $m=1$.
At zero temperature, the free energy density is (cf.\ Eq.~\eqref{FE-def})
\begin{equation}
	F = \frac{1}{F_2}\frac{\Delta^2}{2} + (E - \mu n).
\end{equation}
Here $E$ is the energy density
\begin{multline}
	E = \int_{-4t}^\mu\d\eps\,\eps\D(\eps) \\
		= -(4t)^2 N_0 \Re\left\{ \sqrt{\textstyle 1-(\frac{\mu}{4t})^2}\,
		E\left(1 - \frac{\Delta^2-(\frac{\mu}{2})^2}{(2t)^2-(\frac{\mu}{2})^2}\right)
		\right\},
\end{multline}
where $E(m)$ is the Complete Elliptic Integral of the Second
Kind.\cite{ASellint} And $n$ is the electron density
\begin{multline}
	n = \int_{-4t}^\mu\d\eps\,\D(\eps) \approx \frac{1}{2} + (2tN_0)\mu \\
		{}+ (2tN_0) \left[
			(\Delta-\frac{\mu}{2})
				\ln\frac14\left|\Delta-\frac{\mu}{2}\right|
			- (\mu\to-\mu) \right].
\end{multline}
This expression is a leading order expansion in
$(\Delta\pm\frac{\mu}{2})$. Here, for brevity, $\mu$ and $\Delta$
are in units of $2t$.

Expanding the energy density to the same order, the combined free energy
(neglecting terms independent of $\Delta$) is given in
Eq.~\eqref{FE-expr}. Equating to zero the first derivative of the free
energy with respect to $\Delta$ gives an equation for its local
extrema. Nontrivial minima are easily found at $\mu=0$. In
Eq.~\eqref{Delta-eql} location of these minima is given to quadratic
order in $\mu$. Using this expression we can find the chemical potential
$\mu_c$ (Eq.~\eqref{mu-critical}) at which they become global minima.
Equating to zero the second derivative of the free energy with respect
to $\Delta$, we find the chemical potential $\mu_*$
(Eq.~\eqref{mu-stability}) at which the extremum at $\Delta=0$ changes
from a local minimum to a maximum.

\raggedright
\bibliography{nematic_mfa}

\end{document}